\documentclass[letterpaper]{article}

\usepackage[T1]{fontenc}

\usepackage{geometry}
\usepackage{setspace}
\DeclareMathAlphabet{\mathdutchcal}{U}{dutchcal}{m}{n}
\SetMathAlphabet{\mathdutchcal}{bold}{U}{dutchcal}{b}{n}
\DeclareMathAlphabet{\mathdutchbcal}{U}{dutchcal}{b}{n}
\usepackage[articletitle=true]{achemso}

\usepackage{graphicx}
\usepackage{float}
\newfloat{scheme}{htbp}{los}
\floatname{scheme}{Scheme}
\floatname{chart}{Chart}
\newfloat{graph}{htbp}{loh}

\usepackage{chemformula} % Formulas using \ch{}
\usepackage[version = 4]{mhchem} % Formulas using \ce{}

\usepackage{bm}

\usepackage{xcolor}

\usepackage{authblk}
\author[1]{Toby Kay*}
\author[1]{Serafim Kalliadasis*}
\affil[1]{Complex Multiscale Systems Group, Department of Chemical Engineering,
Imperial College, London SW7 2AZ, United Kingdom}

\title{Orientable Surfactants on Thin Liquid Films: A Dynamic Density-Functional Theory Approach}
\date{*Email: t.kay@imperial.ac.uk, s.kalliadasis@imperial.ac.uk}

\begin{document}

\maketitle

\begin{abstract}
  Thin liquid films are ubiquitous across many natural and engineering
  systems, including films which are laden with surface active molecules,
  i.e. surfactants. The presence of surfactants may have a destabilising
  effect on the film owing to their influence on surface tension, the
  so-called Marangoni effect, which in turn can induce flows in the film.
  Classical thin-film models for surfactant-laden films lead to
  paradigmatic gradient dynamics equations governing the film height and
  surfactant concentration and have been widely studied. However, in all
  these works, which are based on fluid dynamics or nonequilibrium
  thermodynamics, the shape of surfactants is neglected, and they have been
  treated as symmetric point-like particles. In general, this is a drastic
  oversimplification, as surfactants are amphiphilic with a polar head-tail
  structure. To account for this effect we use elements from the
  statistical mechanics of classical fluids, namely
  density-functional theory (DFT), and its dynamic extension (DDFT).
  Starting from DDFT and under the long-wave approximation, we derive the
  pertinent thin-film equations with the surfactants treated as polar
  uniaxial particles. These are equations which govern the film height, as
  well as the surfactant concentration and polarisation field. They
  preserve the gradient dynamics form by appropriately defining the free
  energy, which contains the usual interfacial contributions, as well as
  further contributions from the polarisation field. In doing so, we
  uncover a novel form of a generalised surface tension that is dependent
  on the surfactant polarisation as well as concentration, and show that it
  arises in a thermodynamically consistent way.
\end{abstract}

\section*{Keywords}

Dynamic Density-Functional Theory, Thin-Liquid Films, Orientable Surfactants

%%%%%%%%%%%%%%%%%%%%%%%%%%%%%%%%%%%%%%%%%%%%%%%%%%%%%%%%%%%%%%%%%%%%%
%% Start the main part of the manuscript here.
%%%%%%%%%%%%%%%%%%%%%%%%%%%%%%%%%%%%%%%%%%%%%%%%%%%%%%%%%%%%%%%%%%%%%
\section{Introduction}

Thin-liquid films with surface active agents, so-called surfactant-laden
films, are prevalent in technology and nature; from industrial and commercial
applications with foams, emulsions~\cite{behera2014foaming} and dip
coatings~\cite{wilczek2015modelling}, to biological and physiological
settings such as tear films in the eye~\cite{bhamla2014influence} and
surfactant layers in the alveoli of the lung~\cite{hermans2015lung}.
Surfactants are known to profoundly affect interfacial flows. The so-called
solutocapillary Marangoni effect, is the appearance of flow, or the altering
of an existing flow, due to surface tension non-uniformity, i.e. when the
surface tension varies along an interface with surfactant
concentration~\cite{sternling1959interfacial,scriven1960marangoni}. This
effect can be seen in the well-known phenomena of the ``tears of
wine"~\cite{thomson1855xlii}. Not surprisingly, surfactant-laden films have
received considerable attention over the years.

Typically, their dynamics is governed by two coupled evolution equations for
the height of the film and the surfactant concentration. Two main approaches
to deriving these have been utilised. The first comes from fluid dynamics,
where one starts with the Navier-Stokes equations of motion and associated
wall and free-surface boundary conditions, together with a surface
advection-diffusion equation for the surfactant
concentration~\cite{stone1990simple}, generalised in
Refs.\citenum{Wongetal1996,Antonio2008a}. One then applies the long-wave or
lubrication approximation~\cite{Kalliadasis2012falling} to obtain a set of
two coupled equations for the film height and surfactant, e.g.
Refs.\citenum{oron1997long,Antonio2008b,kopf2009thin}. Alternatively, one may
follow a nonequilibrium thermodynamic approach and construct the set of
coupled equations in terms of an interfacial free-energy functional,
highlighting the gradient dynamics form of the equations, and a mobility
matrix~\cite{thiele2012thermodynamically,thiele2016gradient,voss2024gradient}
(in the same spirit with diffuse-interface methods). With a suitable choice
for the free-energy functional and mobility matrix both approaches lead to
equivalent sets of equations.

Surface-active agents are amphiphilic molecules with a head-tail structure,
where the head and tail are typically hydrophilic and hydrophobic,
respectively, enabling interaction with both polar and non-polar solvents.
This structure gives them a tendency to orient in specific ways on the
surface of a liquid film. In previous works the shape of the surfactant
molecule has been neglected, and surfactant molecules are instead treated as
symmetric point-like particles. This may be a valid approximation in the
dilute regime, where surfactant molecules may be treated as an ideal gas.
However, this is no longer valid for higher surfactant concentrations as the
interactions between the surfactant molecules and the film itself become
increasingly important and surfactant orientation has a significant effect on
these interactions. In the case of a flat-interface two-phase fluid system
with soluble surfactants, it has recently been
shown\cite{hardy2026kinetic,hardy2026surfactant} that the orientation of the
surfactant has a significant effect on the interfacial dynamics, specifically
the orientation induced altering of the effective interface tension. This
implies that for free surface films the orientation of the surfactants will
have an impact on the Marangoni stress and, consequently, on the dynamics of
the film height and surfactant concentration.

In the present study we explicitly account for the shape of surfactants,
specifically their uniaxial polar structure, a significant departure from
previous works. To do so we turn to statistical mechanics of classical
fluids, namely density functional theory (DFT) and its dynamic extension
(DDFT). We start by introducing the DDFT for uniaxial colloidal particles in
a fluid and extend this to when the particles are confined to the surface of
a film, thus obtaining a microscopic derivation of the pertinent surfactant
transport equation. Building on previous surfactant-laden thin-film
derivations, we apply the long-wave approximation and obtain a coupled pair
of evolution equations governing the height of the thin film and the joint
surfactant position and orientation density. Defining the surfactant
concentration and polarisation as moments of this density, we find the
governing equations of these quantities which are coupled to the film-height
equation. To close this set of equations, we define the free-energy
functional in terms of the usual interfacial contributions as well as extra
terms arising from the polarisation of the surfactants. Finally, by defining
a generalised surface tension in terms of the local grand-potential density,
we are able to write the set of thin-film equations in a gradient dynamics
form in terms of the free-energy functional.

%\section{Methods \& Results}

%\subsection{Surface DDFT for Uniaxial Particles}

\section{Surface DDFT for Uniaxial Particles}

\begin{figure}
  \centering
  \includegraphics[width=0.8\textwidth]{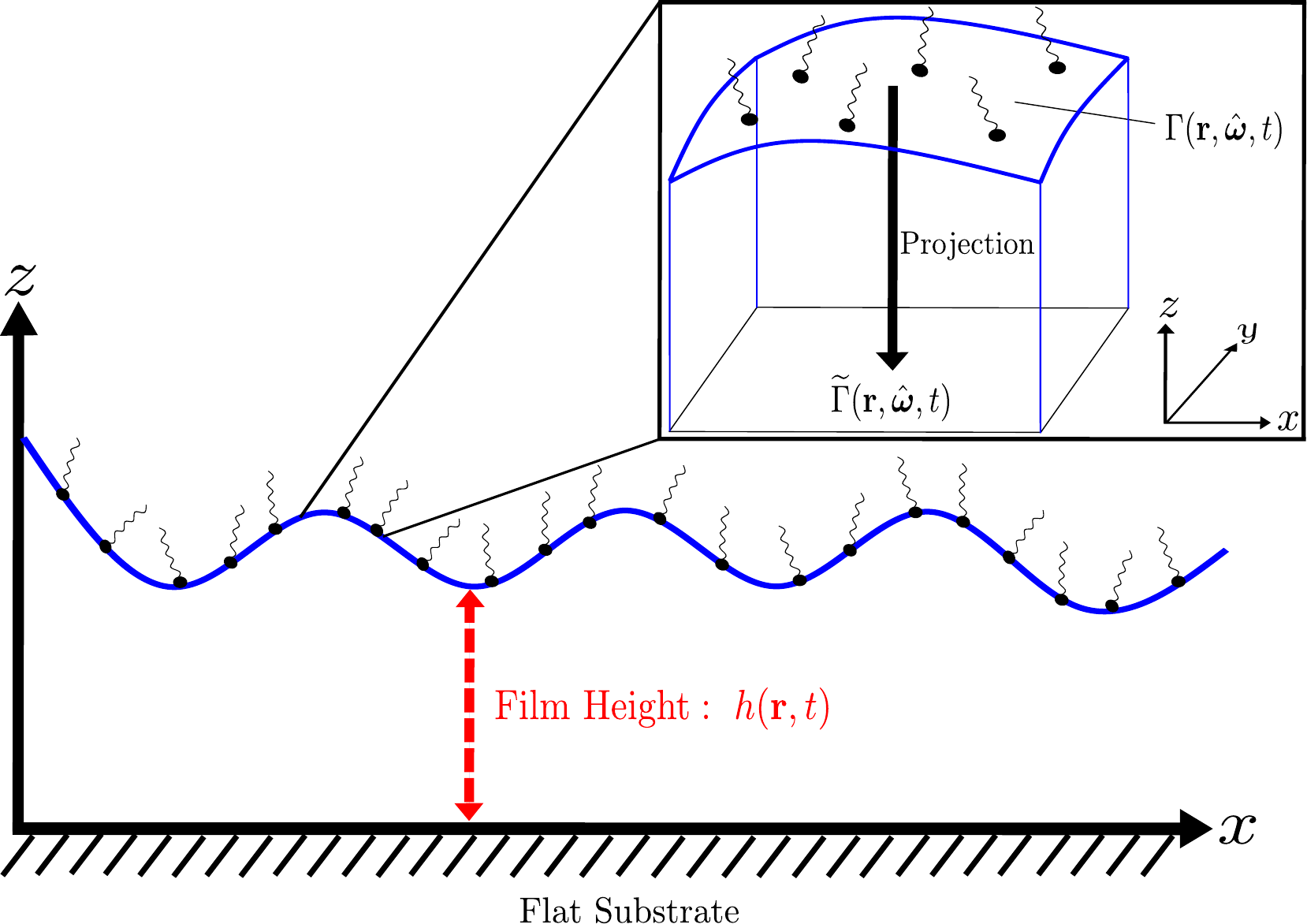}
  \caption{Sketch of the profile geometry for a thin liquid film on a planar horizontal substrate.}
  \label{fig:thin_film}
\end{figure}

Figure \ref{fig:thin_film} depicts the physical setting. A coordinate system
$(x,y,z)$ is chosen so that $x$ and $y$ are the streamwise and spanwise
coordinates, respectively, and $z$ is the outward-pointing coordinate normal
to the wall. The wall is then located at $z=0$ and the film free surface at
$z = h({\mathbf{r}},t)$, where ${\mathbf{r}} = (x ~~ y)^\top$. The film is at
a constant temperature $T$, and its free surface is laden with surfactant
molecules. These surface-active agents are supposed to be of a generic
amphiphilic head-tail structure, such that they may be treated as being polar
uniaxial particles. We will treat them as being insoluble but they diffuse on
the film surface and also assume that their surface concentration is high
enough that molecular interactions between the surfactant molecules cannot be
neglected.

First, we introduce the DDFT for polar uniaxial colloidal particles that
undergo Brownian motion in a generic three-dimensional, $d=3$,
fluid~\cite{rex2007dynamical,wittkowski2011dynamical},
\begin{align}\label{eq:DDFT}
  \partial_t \rho(\bm{\mathdutchcal{r}},\hat{\bm{\omega}},t)= \bm{\nabla}_{\bm{\mathdutchcal{r}}} \cdot \mathbf{D}(\hat{\bm{\omega}}) \cdot \left[\rho(\bm{\mathdutchcal{r}},\hat{\bm{\omega}},t) \bm{\nabla}_{\bm{\mathdutchcal{r}}} \frac{\delta \mathcal{F}[\rho]}{\delta \rho(\bm{\mathdutchcal{r}},\hat{\bm{\omega}},t)} \right] +\mathcal{D}_\text{rot} \hat{\bm{\omega}} \times \bm{\nabla_{\hat{\bm{\omega}}}} \cdot \left[\rho(\bm{\mathdutchcal{r}},\hat{\bm{\omega}},t) \hat{\bm{\omega}} \times \bm{\nabla_{\hat{\bm{\omega}}}} \frac{\delta \mathcal{F}[\rho]}{\delta \rho(\bm{\mathdutchcal{r}},\hat{\bm{\omega}},t)}\right]
\end{align}
where $\bm{\mathdutchcal{r}}=(x ~~ y ~~ z)^\top$ is the $d=3$ positional
coordinate and $\hat{\bm{\omega}}$ is a unit vector giving the particle
orientation pointing from tail to head, such that in the case of apolar
particles, $\hat{\bm{\omega}} \equiv -\hat{\bm{\omega}}$.
$\bm{\nabla}_{\bm{\mathdutchcal{r}}}$ is the Cartesian gradient operator and
$\bm{\nabla}_{\hat{\bm{\omega}}}$ is the gradient on the unit sphere, such
that $\hat{\bm{\omega}} \times \bm{\nabla_{\hat{\bm{\omega}}}}$ is the
rotation operator~\cite{rex2007dynamical}. $\mathcal{D}_\text{rot}$ is the
rotational diffusion coefficient and the translational diffusion tensor for
uniaxial particles is given by\cite{rex2007dynamical}
\begin{equation}\label{eq:diffusion}
  \mathbf{D}(\hat{\bm{\omega}})= D_\parallel \hat{\bm{\omega}} \otimes \hat{\bm{\omega}} + D_\perp \left(\mathbf{I}-\hat{\bm{\omega}} \otimes \hat{\bm{\omega}}\right),
\end{equation}
where $\otimes$ is the dyadic product, and $D_\parallel$ and $D_\perp$ are
the diffusion coefficients parallel and perpendicular to the axis of the
particle, respectively, and $\mathbf{I}$ is the identity matrix.
$\rho(\bm{\mathdutchcal{r}},\hat{\bm{\omega}},t)$ describes the one-body
position and orientation distribution of the colloidal particles
(corresponding to density in fluid dynamics). $\mathcal{F}[\rho]$ is the
free-energy functional, which for now we take it to be just a functional of
the distribution $\rho(\bm{\mathdutchcal{r}},\hat{\bm{\omega}},t)$, which is
of the standard
form~\cite{evans1979nature,goddard2012general,goddard2013unification},
$\mathcal{F}=\mathcal{F}_\text{id}+\mathcal{F}_\text{exc}+\mathcal{F}_\text{ext}$,
where we have contributions from the ideal gas, excess and external free
energies, respectively. At equilibrium the ideal and external free-energy
functionals are given, respectively, by~\cite{rex2007dynamical},
\begin{subequations}\label{eq:ideal_ext_free_energy}
  \begin{align}
    \mathcal{F}_\text{id}[\rho_\text{eq}]&=k_B T\int d\bm{\mathdutchcal{r}}\oint d\hat{\bm{\omega}}\rho_\text{eq}(\bm{\mathdutchcal{r}},\hat{\bm{\omega}}) \left[\ln\mathcal{V}\rho_\text{eq}(\bm{\mathdutchcal{r}},\hat{\bm{\omega}})-1\right],\\
    \mathcal{F}_\text{ext}[\rho_\text{eq}]&=\int d\bm{\mathdutchcal{r}}\oint d\hat{\bm{\omega}}\rho_\text{eq}(\bm{\mathdutchcal{r}},\hat{\bm{\omega}}) V_\text{ext}(\bm{\mathdutchcal{r}},\hat{\bm{\omega}}),
  \end{align}
\end{subequations}
where $\mathcal{V}$ is the thermal volume of the particle, $V_\text{ext}$ is
the external potential, and $\oint d\hat{\bm{\omega}}$ is the integral over
the unit sphere. For the excess free energy we have the
relations\cite{rex2007dynamical},
\begin{subequations}\label{eq:excess_free_energy}
  \begin{align}
    \rho_\text{eq}(\bm{\mathdutchcal{r}},\hat{\bm{\omega}})\bm{\nabla}_{\bm{\mathdutchcal{r}}} \frac{\delta \mathcal{F}_\text{exc}[\rho_\text{eq}]}{\delta \rho_\text{eq}(\bm{\mathdutchcal{r}},\hat{\bm{\omega}})}&=\int d\bm{\mathdutchcal{r}}' \oint d\hat{\bm{\omega}}'\rho_\text{eq}^{(2)}(\bm{\mathdutchcal{r}},\bm{\mathdutchcal{r}}',\hat{\bm{\omega}},\hat{\bm{\omega}}')\bm{\nabla}_{\bm{\mathdutchcal{r}}}U(\bm{\mathdutchcal{r}},\bm{\mathdutchcal{r}}',\hat{\bm{\omega}},\hat{\bm{\omega}}'),\\
    \rho_\text{eq}(\bm{\mathdutchcal{r}},\hat{\bm{\omega}})\hat{\bm{\omega}}\times\bm{\nabla}_{\hat{\bm{\omega}}} \frac{\delta \mathcal{F}_\text{exc}[\rho_\text{eq}]}{\delta \rho_\text{eq}(\bm{\mathdutchcal{r}},\hat{\bm{\omega}})}&=\int d\bm{\mathdutchcal{r}}'\oint d\hat{\bm{\omega}}'\rho_\text{eq}^{(2)}(\bm{\mathdutchcal{r}},\bm{\mathdutchcal{r}}',\hat{\bm{\omega}},\hat{\bm{\omega}}') \hat{\bm{\omega}}\times\bm{\nabla}_{\hat{\bm{\omega}}}U(\bm{\mathdutchcal{r}},\bm{\mathdutchcal{r}}',\hat{\bm{\omega}},\hat{\bm{\omega}}'),
  \end{align}
\end{subequations}
where
$\rho_\text{eq}^{(2)}(\bm{\mathdutchcal{r}},\bm{\mathdutchcal{r}}',\hat{\bm{\omega}},\hat{\bm{\omega}}')$
is the joint equilibrium density and
$U(\bm{\mathdutchcal{r}},\bm{\mathdutchcal{r}}',\hat{\bm{\omega}},\hat{\bm{\omega}}')$
is the interaction potential. Although we are interested in
out-of-equilibrium systems we make the adiabatic
approximation\cite{marconi1999dynamic,marconi2000dynamic,goddard2012general,goddard2013unification},
by which we approximate the higher-body correlations by those of an
equilibrium fluid with the same density; this is effectively an equilibrium
sum rule, by which we assume the system is sufficiently close to equilibrium
and evolves slowly enough that the ideal gas and external free-energy
functionals in Eq. (\ref{eq:ideal_ext_free_energy}) and functional
derivatives of the excess free-energy functional in Eq.
(\ref{eq:excess_free_energy}), may be adequately used in the time-dependent
system of interest. It should be noted though, that, in general, the
adiabatic approximation is uncontrolled and that super-adiabatic effects are
generally present in out-of-equilibrium systems as detailed in Refs.
\citenum{schmidt2022power,de2023perspective}.

As our focus is the motion of surfactants on the surface of the film we need
to adapt Eq. (\ref{eq:DDFT}) to account for lower dimensionality and the
influence of the film dynamics. Related approaches include
(lower-dimensional) DDFT applied to confined
geometries~\cite{penna2003dynamic,archer2005dynamical,goddard2016dynamical,abdoli2026dynamical},
as well as studies on transport of anisotropic colloidal particles at
interfaces\cite{anjali2018shape,yan2022anisotropic,rahman2023rough,kumar2024shape}.
In our case, we consider the density to be describing the motion of polar
uniaxial colloidal particles constrained to the film surface, defined by the
film height $z=h(\mathbf{r},t)$. It is usual to define the joint surfactant
concentration and orientation density as the dimensionless quantity,
$\Gamma(\mathbf{r},\hat{\bm{\omega}},t)=\ell^2
\rho(\mathbf{r},\hat{\bm{\omega}},t)$, where $\ell^2$ is the molecular length
scale, i.e. the surface area per surfactant molecule when there is maximum
packing on the surface\cite{thiele2012thermodynamically}. From Eq.
(\ref{eq:DDFT}) and Ref.\citenum{stone1990simple} we are able to formulate
the following transport equation governing the surfactant concentration on
the free surface, (details are given in Appendix A),
\begin{equation}\label{eq:surface_DDFT}
  \partial_t \Gamma=-\bm{\nabla}_s \cdot (\Gamma \mathbf{v}_s) - \Gamma \left(\bm{\nabla}_s \cdot \hat{\mathbf{n}} \right) (\mathbf{v}\cdot \hat{\mathbf{n}}) + \bm{\nabla}_s \cdot \mathbf{D}_s \cdot \left[ \Gamma \bm{\nabla}_s \frac{\delta \mathcal{F}}{\delta \Gamma} \right] +  \mathcal{D}_\text{rot} \hat{\bm{\omega}} \times \bm{\nabla}_{\hat{\bm{\omega}}} \cdot \left[\Gamma \hat{\bm{\omega}} \times \bm{\nabla}_{\hat{\bm{\omega}}} \frac{\delta \mathcal{F}}{\delta \Gamma} \right],
\end{equation}
where we have implicitly included the rescaling by $\ell^2$. Here
$\mathbf{v}$ represents the velocity of the film interface and
$\mathbf{v}_s=(\mathbf{I}-\hat{\mathbf{n}}\otimes \hat{\mathbf{n}})\cdot
\mathbf{v}$ is the velocity tangential to the film surface, where
\begin{equation}\label{eq:normal}
    \hat{\mathbf{n}}=\frac{(-\partial_x h ~~ -\partial_y h ~~ 1)^\top}{\sqrt{1+|\bm{\nabla} h|^2}},
\end{equation}
is the unit normal (outward-pointing) to the surface. Then
$\bm{\nabla}_s=(\mathbf{I}-\hat{\mathbf{n}}\otimes \hat{\mathbf{n}})\cdot
\bm{\nabla}_{\bm{\mathdutchcal{r}}}$ is the surface gradient and
$\mathbf{D}_s=(\mathbf{I}-\hat{\mathbf{n}}\otimes \hat{\mathbf{n}}) \cdot
\mathbf{D} \cdot (\mathbf{I}-\hat{\mathbf{n}}\otimes \hat{\mathbf{n}})$ is
the translational diffusion tensor projected onto the surface. In Eq.
(\ref{eq:surface_DDFT}) we leave the rotation term unmodified as compared to
Eq. (\ref{eq:DDFT}), as we allow the surfactant molecules to rotate freely in
three dimensions.

There is a rather subtle point associated with the surfactant transport
equation in Eq. (\ref{eq:surface_DDFT}), and more general, a transport
equation for an interfacial quantity, such as mass, mole, or charge density,
$\Gamma$. The partial time derivative, $\partial \Gamma/\partial t$, has to
be defined with care. Specifically, in Ref.~\citenum{Antonio2008a}, a
generalised transport equation for an interfacial entity was derived. In this
equation, $\partial \Gamma/\partial t$ is interpreted as the partial time
derivative of $\Gamma$ with respect to a local frame of reference moving with
the velocity normal to the surface, $\mathbf{v}_{\rm n}$: it is the ``normal
time derivative", $(\partial \Gamma/\partial t)_{\rm n}$. This transport
equation agrees with the one derived in Ref.~\citenum{stone1990simple}
provided that the time derivative in Ref.~\citenum{stone1990simple} is
\text{not} interpreted as the usual partial time derivative, i.e. the
derivative obtained by differentiating with respect to time for a fixed point
in space, hence the time derivative in the laboratory frame, but as
$(\partial \Gamma/\partial t)_{\rm n}$. The same conclusion was reached in
Ref.~\citenum{Wongetal1996} but with a more involved derivation to that in
Ref.~\citenum{Antonio2008a}.

%\subsection{Long-Wave Expansion}

\section{Long-Wave Expansion}

Equation (\ref{eq:surface_DDFT}) is quite intricate, but it can be greatly
simplified in the framework of the long-wave
approximation~\cite{Kalliadasis2012falling}. The approximation is widely
known and has been used successfully for a wide range of thin-flow settings
in the presence of various levels of complexity, including
thermo-/solutocapillary Marangoni effects and substrate topography, e.g.
Refs.\citenum{oron1997long,Catherine2000}. It relies on the assumption that
waves on the surface of the film, are typically long compared to the film
thickness, or equivalently, deformations of the free surface are weak. They
are also slow in time. Equivalently, the evolution of thin films is dominated
by long-wave modes~\cite{Kalliadasis2012falling}. This is because the
viscosity of the fluid ensures a strong coherence of the flow across the
film. As a consequence, variations of the interface are slow in time and
space.

These fortuitous characteristics, inherent to thin films, enable a drastic
reduction of the complexity of the governing equations and boundary
conditions, and to obtain systems of simplified model equations. To express
the smallness of the interfacial slope, a small parameter $\varepsilon \sim
|\bm{\nabla}h|\ll 1$ is introduced, such that $\partial_{x,y,t} \sim
\varepsilon$. The film can then be seen as sufficiently flat, i.e.
$\hat{\mathbf{n}}\approx (0 ~~ 0 ~~ 1)^\top$ to leading order in
$\varepsilon$, and evolving sufficiently slowly. Under this approximation, we
have that the interfacial velocity $\mathbf{v}$, is dominated by its
tangential components. Therefore, the normal component is small in
comparison, which leads to the term $\Gamma \left(\bm{\nabla}_s \cdot
\hat{\mathbf{n}} \right) (\mathbf{v}\cdot \hat{\mathbf{n}})$ in Eq.
(\ref{eq:surface_DDFT}) only contributing at higher orders and can thus be
neglected. Similarly, at leading order the surface gradient can be replaced
by the $d=2$ Cartesian gradient, $\bm{\nabla}_\mathbf{r}$, leading to the
long-wave form of Eq. (\ref{eq:surface_DDFT}),
\begin{equation}\label{eq:long_wave_ddft}
  \partial_t \Gamma=-\bm{\nabla}_\mathbf{r} \cdot (\Gamma \mathbf{v}^{xy}_s) + \bm{\nabla}_\mathbf{r} \cdot \mathbf{D}^{xy} \cdot \left[ \Gamma \bm{\nabla}_\mathbf{r} \frac{\delta
  \mathcal{F}_I}{\delta \Gamma} \right] +  \mathcal{D}_\text{rot} \hat{\bm{\omega}} \times \bm{\nabla}_{\hat{\bm{\omega}}} \cdot \left[\Gamma \hat{\bm{\omega}} \times \bm{\nabla}_{\hat{\bm{\omega}}}
  \frac{\delta \mathcal{F}_I}{\delta \Gamma} \right],
\end{equation}
where $\mathbf{v}^{xy}=(v_x ~~ v_y)^\top$ is the velocity vector of the
interface parallel to the substrate and
$\mathbf{v}_s^{xy}=\mathbf{v}^{xy}(h)$ is the corresponding velocity at the
interface. $\mathbf{D}^{xy}$ is the ($2\times 2$) diffusion tensor with only
the $x$- and $y$-components retained. $\mathcal{F}_I$ is the interfacial free
energy which will have additional contributions due to surface dynamics as
compared to the general free energy, $\mathcal{F}$, introduced in the
previous section.

Now, we calculate the velocity, $\mathbf{v}^{xy}$ that appears in Eq.
(\ref{eq:long_wave_ddft}). This is a standard calculation using the long-wave
expansion, where one can show that $\mathbf{v}^{xy}$ satisfies the Stokes
equation, e.g.
Refs.\citenum{oron1997long,Kalliadasis2012falling,thiele2016gradient},
\begin{equation}\label{eq:Stokes}
  \eta \partial_z^2 \mathbf{v}^{xy}(z)=\bm{\nabla}_\mathbf{r}p,
\end{equation}
where $\eta$ is the (bulk) viscosity and $p$ is the pressure. Equation
(\ref{eq:Stokes}) may be trivially solved by a double integration over $z$, and
employing the no-slip boundary condition on the substrate,
$\mathbf{v}^{xy}(0)=0$, as well as the leading-order stress balance condition
on the interface, $\eta \partial_z
\mathbf{v}^{xy}(h)=\bm{\nabla}_\mathbf{r}\gamma$, where $\gamma$ is the
interfacial tension. One then obtains the following expression for the
velocity,
\begin{equation}\label{eq:velocity}
  \mathbf{v}^{xy}(z)=-\frac{z}{\eta}\left[\left(h-\frac{z}{2}\right)\bm{\nabla}_\mathbf{r}p - \bm{\nabla}_\mathbf{r} \gamma \right],
\end{equation}
such that the surface velocity is given by,
\begin{equation}\label{eq:surface_velocity}
  \mathbf{v}_s^{xy}=-\frac{h^2}{2\eta} \bm{\nabla}_\mathbf{r}p + \frac{h}{\eta} \bm{\nabla}_\mathbf{r} \gamma.
\end{equation}

From Eq. (\ref{eq:velocity}) we are able to write the governing equation for
the height of the film, $h(\mathbf{r},t)$, via the conservation law or
continuity equation,
\begin{equation}
  \partial_t h = -\bm{\nabla}_\mathbf{r} \cdot \int_0^h dz
\mathbf{v}^{xy}(z),
\end{equation}
which gives the well-known form, e.g.
Refs.\citenum{Antonio2008b,thiele2012thermodynamically,wilczek2015modelling,thiele2016gradient,voss2024gradient}
\begin{equation}\label{eq:thin_film}
  \partial_t h = \bm{\nabla}_\mathbf{r} \cdot \left[ \frac{h^3}{3\eta} \bm{\nabla}_\mathbf{r}p - \frac{h^2}{2\eta} \bm{\nabla}_\mathbf{r} \gamma \right].
\end{equation}
We may relate the pressure in Eq. (\ref{eq:thin_film}) to the free-energy
functional by explicitly including the dependence of the free energy on the
height, and concentration $\Gamma$, as $\mathcal{F}_I[h,\Gamma]$, such
that~\cite{thiele2016gradient}
\begin{equation}
  p(\mathbf{r},t)=\frac{\delta \mathcal{F}_I[h,\Gamma]}{\delta h(\mathbf{r},t)}.
\end{equation}
However, the consequence of this is that the height, $h$, and concentration,
$\Gamma$, are not independent as $\Gamma$ is the concentration on the curved
surface, defined by $h$~\cite{wilczek2015modelling}. To remedy this, we take
$\widetilde{\Gamma}=\xi \Gamma$, where
$\xi=\sqrt{1+|\bm{\nabla}_\mathbf{r}h|^2}$ is the metric factor of the film
surface, such that $\widetilde{\Gamma}$ is the surfactant concentration
projected onto the flat substrate plane, as illustrated in Fig.
\ref{fig:thin_film}. Therefore, we write the free-energy functional as
$\mathcal{F}_I[h,\widetilde{\Gamma}/\xi]$, and take the variations with
respect to $\widetilde{\Gamma}$. Similarly, we may use Eq.
(\ref{eq:surface_velocity}) in Eq. (\ref{eq:long_wave_ddft}) to write the
governing equation for $\Gamma(\mathbf{r},\hat{\bm{\omega}},t)$,
\begin{equation}\label{eq:joint_concentration_thin_film}
  \partial_t \Gamma=\bm{\nabla}_\mathbf{r} \cdot \left[ \frac{h^2 \Gamma}{2\eta} \bm{\nabla}_\mathbf{r}
  \frac{\delta \mathcal{F}_I}{\delta h} - \frac{h \Gamma}{\eta} \bm{\nabla}_\mathbf{r} \gamma \right] + \bm{\nabla}_\mathbf{r} \cdot \mathbf{D}^{xy} \cdot \left[ \Gamma \bm{\nabla}_\mathbf{r}
  \frac{\delta \mathcal{F}_I}{\delta \widetilde{\Gamma}} \right] +  \mathcal{D}_\text{rot} \hat{\bm{\omega}} \times \bm{\nabla}_{\hat{\bm{\omega}}} \cdot \left[\Gamma \hat{\bm{\omega}} \times \bm{\nabla}_{\hat{\bm{\omega}}}
  \frac{\delta \mathcal{F}_I}{\delta \widetilde{\Gamma}} \right],
\end{equation}
which is coupled to the governing equation for the film height, Eq. (\ref{eq:thin_film}).

To further simplify Eq. (\ref{eq:joint_concentration_thin_film}), we must
obtain the form of the functional, $\mathcal{F}_I$, which will include
interfacial contributions as well contributions from the dynamics of the
surfactants, as in Eqs. (\ref{eq:ideal_ext_free_energy}) and
(\ref{eq:excess_free_energy}). Due to the inclusion of the surfactant
orientation in $\Gamma$, we may also add in the free-energy functional
orientation-derived effects on the interface. To do so, we coarse-grain by
considering the first two moments of $\Gamma$ with respect to
$\hat{\bm{\omega}}$. First we have the zeroth-moment, the marginal of
$\Gamma$ over $\hat{\bm{\omega}}$, as the surface concentration,
$c(\mathbf{r},t)$, where
\begin{equation}\label{eq:concentration}
  c(\mathbf{r},t)=\oint d\hat{\bm{\omega}} \Gamma(\mathbf{r},\hat{\bm{\omega}},t),
\end{equation}
and the first-moment as the expectation over $\hat{\bm{\omega}}$ as,
\begin{equation}\label{eq:polarisation}
  \mathbf{P}(\mathbf{r},t)=\oint d\hat{\bm{\omega}} \ \hat{\bm{\omega}} \Gamma(\mathbf{r},\hat{\bm{\omega}},t),
\end{equation}
which defines the polarisation vector field of the surfactants. The
polarisation gives the local average orientation of the surfactant molecules,
such that it measures how strongly and in which direction the surfactant
molecules are pointing. Clearly, for apolar molecules, $\mathbf{P}$, will
vanish, but surfactant molecules, with their head-tail structure, are polar,
and $\mathbf{P}$, in general will be non-zero. If we make the assumption that
the orientation-based contributions are wholly defined by the polarisation,
$\mathbf{P}$, we are able to approximate the free-energy functional as,
\begin{equation}\label{eq:free_energy_approx}
  \mathcal{F}_I[h,\Gamma]\approx \mathcal{F}_I[h,c,\mathbf{P}].
\end{equation}
By using Eq. (\ref{eq:free_energy_approx}) in Eq.
(\ref{eq:joint_concentration_thin_film}) and performing the integrations in
Eqs.~(\ref{eq:concentration}) and~(\ref{eq:polarisation}) we are able to
derive thin-film equations for the surfactant concentration and polarisation.

\section{Thin-Film Equations for Surfactant Concentration and Polarisation}

To perform the coarse-graining of Eq.
(\ref{eq:joint_concentration_thin_film}) to find the governing equations of
$c$ and $\mathbf{P}$, we must first express the variational derivatives of
the free energy functional in terms of $\widetilde{c}=\xi c$ and
$\widetilde{\mathbf{P}}=\xi \mathbf{P}$. To do so, we use the fact that
$\widetilde{c}$ and $\widetilde{\mathbf{P}}$ are functionals of
$\widetilde{\Gamma}$, which means we may calculate the variational
derivatives using the functional chain rule, such that,
\begin{equation}
  \frac{\delta \mathcal{F}_I}{\delta \widetilde{\Gamma}(\mathbf{r},\hat{\bm{\omega}},t)}=\int d\mathbf{r}'
  \left[\frac{\delta \mathcal{F}_I}{\delta \widetilde{c}(\mathbf{r}',t)}\frac{\delta \widetilde{c}(\mathbf{r}',t)}{\delta \widetilde{\Gamma}(\mathbf{r},\hat{\bm{\omega}},t)} +
  \frac{\delta \mathcal{F}_I}{\delta \widetilde{\mathbf{P}}(\mathbf{r}',t)} \cdot \frac{\delta \widetilde{\mathbf{P}}(\mathbf{r}',t)}{\delta \widetilde{\Gamma}(\mathbf{r},\hat{\bm{\omega}},t)} \right].
\end{equation}
From Eqs. (\ref{eq:concentration}) and (\ref{eq:polarisation}), we have that
\begin{subequations}
  \begin{align}
    \frac{\delta \widetilde{c}(\mathbf{r}',t)}{\delta \widetilde{\Gamma}(\mathbf{r},\hat{\bm{\omega}},t)}&= \oint d\hat{\bm{\omega}}' \delta(\mathbf{r}-\mathbf{r}') \delta_\text{rot}(\hat{\bm{\omega}},\hat{\bm{\omega}}')=\delta(\mathbf{r}-\mathbf{r}'),\\
    \frac{\delta \widetilde{\mathbf{P}}(\mathbf{r}',t)}{\delta \widetilde{\Gamma}(\mathbf{r},\hat{\bm{\omega}},t)}&= \oint d\hat{\bm{\omega}}' \ \hat{\bm{\omega}}' \delta(\mathbf{r}-\mathbf{r}') \delta_\text{rot}(\hat{\bm{\omega}},\hat{\bm{\omega}}')=\hat{\bm{\omega}}\delta(\mathbf{r}-\mathbf{r}'),
  \end{align}
\end{subequations}
where $\delta_\text{rot}(\hat{\bm{\omega}},\hat{\bm{\omega}}')$ is the rotational Dirac-$\delta$ such that for an arbitrary function, $g(\hat{\bm{\omega}})$, $\oint d\hat{\bm{\omega}}' g(\hat{\bm{\omega}}') \delta_\text{rot}(\hat{\bm{\omega}},\hat{\bm{\omega}}')=g(\hat{\bm{\omega}})$, and we have used the identity, $\delta g(\hat{\bm{\omega}}')/\delta g(\hat{\bm{\omega}})=\delta_\text{rot}(\hat{\bm{\omega}},\hat{\bm{\omega}}')$. Thus, we obtain,
\begin{equation}\label{eq:chain_rule}
  \frac{\delta \mathcal{F}_I}{\delta \widetilde{\Gamma}(\mathbf{r},\hat{\bm{\omega}},t)}=\frac{\delta \mathcal{F}_I}{\delta \widetilde{c}(\mathbf{r},t)} + \hat{\bm{\omega}} \cdot
  \frac{\delta \mathcal{F}_I}{\delta \widetilde{\mathbf{P}}(\mathbf{r},t)}.
\end{equation}

Now, we are in a position to derive the equations for surfactant
concentration and polarisation. Consider first the equation for the
concentration, $\widetilde{c}(\mathbf{r},t)$, which from Eq.
(\ref{eq:concentration}), and integrating Eq.
(\ref{eq:joint_concentration_thin_film}) over $\hat{\bm{\omega}}$, yields
\begin{align}\label{eq:concentration_thin_film}
  \partial_t c &= \bm{\nabla}_\mathbf{r} \cdot \left[ \frac{h^2 c}{2\eta} \bm{\nabla}_\mathbf{r} \frac{\delta \mathcal{F}_I}{\delta h} - \frac{h}{\eta} c \bm{\nabla}_\mathbf{r} \gamma \right] +\bm{\nabla}_\mathbf{r}\cdot \left[(D_\parallel-D_\perp) \mathbf{Q}^{xy}\cdot \bm{\nabla}_\mathbf{r}
  \frac{\delta \mathcal{F}_I}{\delta \widetilde{c}}\right] \nonumber \\
  &+ \bm{\nabla}_\mathbf{r} \cdot \left[\left(\frac{D_\parallel}{3}+\frac{2D_\perp}{3} \right) c \bm{\nabla}_\mathbf{r}
  \frac{\delta \mathcal{F}_I}{\delta \widetilde{c}}\right] +\bm{\nabla}_\mathbf{r} \cdot \left[D_\perp \left(\mathbf{P} \cdot \bm{\nabla}_\mathbf{r}\right)
  \frac{\delta \mathcal{F}_I}{\delta \widetilde{\mathbf{P}}}\right],
\end{align}
where the rotational term in Eq. (\ref{eq:joint_concentration_thin_film})
integrates to zero and we have assumed moments higher-than-second are small
and can be neglected. This assumption may be made due to the rapid relaxation
of higher angular modes under rotational diffusion and the strong anchoring
due to the amphiphilic nature of the surfactants. This means that these
moments decay faster than the lower moments, and thus, they may be neglected
under the adiabatic approximation. Recall, $D_\perp$ and $D_\parallel$ are
the perpendicular and parallel diffusion coefficients from the diffusion
tensor in Eq. (\ref{eq:diffusion}), respectively. We have also taken the
interfacial tension, $\gamma$, to be independent of $\hat{\bm{\omega}}$
(although it can be dependent on $c$ and $\mathbf{P}$).
$\mathbf{Q}(\mathbf{r},t)$ is a traceless symmetric tensor defined
by\cite{de1993physics,marchetti2013hydrodynamics},
\begin{equation}\label{eq:Q_tensor}
  \mathbf{Q}(\mathbf{r},t)=\oint d\hat{\bm{\omega}} \left(\hat{\bm{\omega}}\otimes \hat{\bm{\omega}}-\frac{1}{3}\mathbf{I}\right) \Gamma(\mathbf{r},\hat{\bm{\omega}},t),
\end{equation}
which describes the local alignment of the particle axes without
distinguishing head from tail and vanishes in the isotropic
case\cite{marchetti2013hydrodynamics}. Thus, $\mathbf{Q}^{xy}$ is the block
matrix of the $x$- and $y$-components of $\mathbf{Q}$.

Similarly, using Eq. (\ref{eq:polarisation}) we may average over Eq.
(\ref{eq:joint_concentration_thin_film}) to obtain the polarisation thin-film
equation (details are given in Appendix B),
\begin{align}\label{eq:polarisation_thin_film}
  \partial_t \mathbf{P} &= \bm{\nabla}_\mathbf{r} \cdot \left[ \frac{h^2 }{2\eta} \mathbf{P} \otimes \bm{\nabla}_\mathbf{r}
  \frac{\delta \mathcal{F}_I}{\delta h} - \frac{h}{\eta} \mathbf{P} \otimes \bm{\nabla}_\mathbf{r} \gamma \right] + \bm{\nabla}_\mathbf{r} \cdot \left[D_\perp \mathbf{P} \otimes \bm{\nabla}_\mathbf{r}
  \frac{\delta \mathcal{F}_I}{\delta \widetilde{c}}\right] + \mathcal{D}_\text{rot} \left(\mathbf{Q}-\frac{2}{3}c \mathbf{I} \right)\cdot
  \frac{\delta \mathcal{F}_I}{\delta \widetilde{\mathbf{P}}},
\end{align}
where again we have neglected all moments higher than the second.
Interestingly, in Eq. (\ref{eq:polarisation_thin_film}) the rotation term
does contribute to the dynamics, and is in a non-conserved form.

Equations (\ref{eq:concentration_thin_film}) and
(\ref{eq:polarisation_thin_film}) along with the equation for the film
height, Eq. (\ref{eq:thin_film}), comprise the set of equations that describe
the dynamics of a thin liquid film with a surfactant-laden interface, where
the surfactants are orientable. However, in their current state, this set of
equations is not closed and hence not particularly useful. The occurrence of
the tensor $\mathbf{Q}$ means we have to supplement our set of equations with
a governing equation for $\mathbf{Q}$. However, this equation will itself
contain higher-order moments, and we will have an infinite hierarchy of
moment equations, which means we still are not able to get closure. In the
case of nematic liquid crystals many closure approximations exist, e.g. Refs.
\citenum{doi1986theory,feng1998closure,wang2013microscopic} and references
therein. In these cases the fluid is apolar and thus the tensor $\mathbf{Q}$
is the lowest-order moment of interest, such that the moment closures are
concerned with higher-order tensors. In our case, the surfactants are
strongly polar, so we are concerned with a moment closure for $\mathbf{Q}$ as
the lower order moment $\mathbf{P}$ is non-zero. Therefore, we assume that
the polarisation is dominant and that $\mathbf{Q}$ relaxes quickly, in
comparison, and thus may be adiabatically slaved to the polarisation
$\mathbf{P}$ (and concentration), through some function $\mathbf{q}$, i.e.
$\mathbf{Q}\approx \mathbf{q}(c,\mathbf{P})$. $\mathbf{q}(c,\mathbf{P})$ may be
approximated by making the argument that the surfactants are strongly aligned
perpendicular to the surface, which seems reasonable for surfactants, due to
them being strongly polar. So, when the surfactants are exactly aligned
perpendicular to the film surface, we may take that $\mathbf{q}$ be written
in terms of the surface normal, $\hat{\mathbf{n}}$, as,
\begin{equation}
  \mathbf{q}=c\left(\hat{\mathbf{n}}\otimes \hat{\mathbf{n}}-\frac{1}{3}\mathbf{I}\right).
\end{equation}
Now, if we take that the polarisation may be written in terms of the surface
normal plus a small perturbation, i.e.
$\mathbf{P}=c\mathbf{\hat{\mathbf{n}}}+\delta \mathbf{P}$, then for $\delta
\mathbf{P}\ll c$, up to leading order we may write $\mathbf{q} =
\mathbf{P}\otimes \mathbf{P}/c-c\mathbf{I}/3+ \mathcal{O}(\delta
\mathbf{P}/c)$. Thus, under the assumption of strong polar alignment of the
surfactants, we may write the leading-order approximation to the tensor
$\mathbf{Q}$, slaved to $c$ and $\mathbf{P}$, as
\begin{equation}
  \mathbf{Q}(\mathbf{r},t)\approx \frac{\mathbf{P}(\mathbf{r},t)\otimes \mathbf{P}(\mathbf{r},t)}{c(\mathbf{r},t)}-\frac{c(\mathbf{r},t)}{3}\mathbf{I}.
\end{equation}

%\subsection{Interfacial Free-Energy Functional}

\section{Interfacial Free-Energy Functional}

In the thin film equations (\ref{eq:thin_film}),
(\ref{eq:concentration_thin_film}) and (\ref{eq:polarisation_thin_film}), the
interfacial free energy is yet to be defined, thus we must express this
functional explicitly in terms of the height, surfactant concentration and
polarisation to be able to carry out the functional derivatives that follow.
The free-energy functional will be made up of different contributions,
arising from different aspects of the system, which in turn depend on the
different observables present, namely, film height, surfactant concentration
and polarisation. The only contribution that depends purely on the height of
the film is from the binding potential~\cite{Peter2016}, $\Phi(h)$, which
describes the interaction between the substrate and the film, and is related
to the so-called disjoining or Derjaguin pressure going back to
Derjaguin~\cite{Derjaguin1986,Derjaguin1987} and Frumkin~\cite{Frumkin1938},
$\Pi=-\Phi'(h)$, valid for parallel or nearly-parallel interfaces (e.g.
liquid-vapor, liquid solid as is the case here); it is the extra pressure
acting on a substrate due to the presence of a thin liquid film, e.g.
Refs.\citenum{de1985wetting,Andreas2014,Peter2021}.

Let us note that one may include concentration dependence in the wetting
potential which will introduce an extra term to the Marangoni
force\cite{thiele2012thermodynamically,thiele2016gradient}. Then, we have the
contribution due to the surfactant molecules on the surface through the local
interfacial free energy\cite{thiele2016gradient}, $f_s(c,\mathbf{P})$, and we
include a term to penalise spatial inhomogeneities in surfactant
concentration which is to account for surfactant phase
transitions\cite{kopf2009thin,wilczek2015modelling}. Similarly, we have the
contributions from the surfactant polarisation which can emerge from the
energy of spontaneous polarisation, elastic
energy\cite{marchetti2013hydrodynamics} and the anchoring of the hydrophilic
heads of the surfactant to the film surface. Additionally, one may include
contributions from the coupling between the different
observables\cite{trinschek2020thin,stegemerten2022symmetry}, however, these
shall be omitted: our aim is to consider a minimal model that captures the
main ingredients of the system by focusing on the most significant effects.
Thus, the interfacial free-energy functional is written as,
\begin{align}\label{eq:free_energy_functional}
  \mathcal{F}_I[h,c,\mathbf{P}]&=\int d\mathbf{r}\Bigg[\Phi(h) + \xi \left(f_s(c,\mathbf{P}) + \frac{\kappa_c}{2} |\bm{\nabla}_s c|^2 \right) \nonumber \\
  &+ \xi\left(\frac{a}{2} \left|\frac{\mathbf{P}}{c}\right|^2 + \frac{b}{4} \left|\frac{\mathbf{P}}{c}\right|^4  + \frac{\kappa_P}{2} \left(\bm{\nabla}_s \frac{\mathbf{P}}{c}\right) : \left(\bm{\nabla}_s \frac{\mathbf{P}}{c}\right) + W \frac{\mathbf{P}}{c}\cdot \hat{\mathbf{n}} \right)  \Bigg].
\end{align}
Here $\kappa_c$ is the interfacial rigidity constant and appears in the term
responsible for penalising concentration gradients, and is also found in the
free-energy associated with the Cahn-Hilliard equation\cite{cahn1958free}.
$a$ and $b$ arise from the spontaneous polarisation term, where in the case
of surfactants this originates from the orientational
entropy\cite{hardy2026surfactant}, such that $a=3k_BTc$, and the quartic term
is present, with $b>0$, for stability to prevent $\mathbf{P}$ growing
uncontrollably and to include saturation
effects\cite{marchetti2013hydrodynamics}. $\kappa_P$ is the Frank elastic
constant\cite{marchetti2013hydrodynamics} which is typically of the
order\cite{klus2014all} $\kappa_P \sim 10^{-12}\textrm{N}$, where the
associated term penalises large gradients in the orientation of the
surfactants, smoothing out the polarisation field. $W$ is the anchoring
strength, where $W>0$ favours the head of the surfactant molecules pointing
into the surface, that is the polarisation field is anti-parallel to the
surface normal, which accounts for the hydrophilic head and hydrophobic tail
of the surfactant\cite{zhang2024surface}, and $W$ is typically of the
order\cite{yesil2018anchoring} $W\sim 10^{-6}\textrm{J}/\textrm{m}^2$. The
local interfacial free energy, $f_s(c,\mathbf{P})$, is given in terms of
different contributions as\cite{thiele2016gradient},
\begin{equation}\label{eq:local_surface_free_energy}
  f_s(c,\mathbf{P})=\gamma_0+\frac{k_B T}{\ell^2}c(\ln c-1) + f_s^\text{exc}(c,\mathbf{P}),
\end{equation}
where $\gamma_0$ is the surface tension in the absence of surfactants, the
second term in Eq. (\ref{eq:local_surface_free_energy}) is the ideal gas
contribution, and $f_s^\text{exc}(c,\mathbf{P})$ is the excess-over-ideal
free energy which accounts for interactions between surfactant molecules (see
Eq. (\ref{eq:excess_free_energy})). Here we assume that the contribution of
the surfactant orientation to the local excess free energy is entirely
through the polarisation field.

From Eq. (\ref{eq:free_energy_functional}) we may compute the functional
derivatives in Eqs. (\ref{eq:thin_film}), (\ref{eq:concentration_thin_film})
and (\ref{eq:polarisation_thin_film}) to fully close this set of equations.
Recalling, we must replace $c$ and $\mathbf{P}$ by $\widetilde{c}/\xi$ and
$\widetilde{\mathbf{P}}/\xi$, respectively, in Eq.
(\ref{eq:free_energy_functional}) to perform the variations. After doing so,
and by taking the leading-order terms in $\varepsilon$, in accordance with
the long-wave approximation, $\varepsilon \sim
|\bm{\nabla}_{\mathbf{r}}h|\ll 1$ such that we take $\xi\approx 1$, we find
the variational derivative with respect to the liquid film height
as\cite{thiele2016gradient},
\begin{align}\label{eq:height_var}
  \frac{\delta \mathcal{F}_I}{\delta h(\mathbf{r},t)}&= \partial_h \Phi(h) -  \bm{\nabla}_\mathbf{r}\cdot \left[\left(f_s(c,\mathbf{P})-c\partial_c f_s(c,\mathbf{P})
  - \mathbf{P}\cdot \bm{\nabla}_\mathbf{P} f_s(c,\mathbf{P}) -\frac{\kappa_c}{2}|\bm{\nabla}_\mathbf{r}c|^2 +\kappa_c c\bm{\nabla}_\mathbf{r}^2 c  \right) \bm{\nabla}_\mathbf{r}h \right]\nonumber \\
  &- \bm{\nabla}_\mathbf{r}\cdot \left[\left(\frac{a}{2} \frac{\left|\mathbf{P}\right|^2}{c^2} + \frac{b}{4} \frac{\left|\mathbf{P}\right|^4}{c^4}  + \frac{\kappa_P}{2} \left(\bm{\nabla}_\mathbf{r} \frac{\mathbf{P}}{c}\right) : \left(\bm{\nabla}_\mathbf{r} \frac{\mathbf{P}}{c}\right) + W \frac{\mathbf{P}}{c} \cdot \hat{\mathbf{n}}\right) \bm{\nabla}_\mathbf{r}h\right],
\end{align}
where under the long-wave approximation we have replaced the surface gradient
with the $d=2$ Cartesian gradient, $\bm{\nabla}_\mathbf{r}$. The variations
with respect to $\widetilde{c}$ and $\widetilde{\mathbf{P}}$ are,
\begin{align}
  \frac{\delta \mathcal{F}_I}{\delta \widetilde{c}(\mathbf{r},t)}&=\partial_c f_s(c,\mathbf{P}) - \kappa_c \bm{\nabla}_\mathbf{r}^2 c - \frac{1}{c}\left(a \frac{\left|\mathbf{P}\right|^2}{c^2} + b \frac{\left|\mathbf{P}\right|^4}{c^4}\right) + \kappa_P \frac{\mathbf{P}}{c^2} \cdot \bm{\nabla}_\mathbf{r}^2 \frac{\mathbf{P}}{c} - W \frac{\mathbf{P}}{c^2}\cdot \hat{\mathbf{n}},\label{eq:concentration_derivative}\\
  \frac{\delta \mathcal{F}_I}{\delta \widetilde{\mathbf{P}}(\mathbf{r},t)}& = \bm{\nabla}_\mathbf{P}f_s(c,\mathbf{P}) + \mathbf{P}\left(\frac{a}{c^2}+b \frac{|\mathbf{P}|^2}{c^4}\right) -\kappa_P\frac{1}{c}\bm{\nabla}_\mathbf{r}^2 \frac{\mathbf{P}}{c} + W \frac{1}{c}\hat{\mathbf{n}}.\label{eq:polarisation_derivative}
\end{align}

Let us now define the generalised interfacial tension, $\gamma$, as
\begin{align}\label{eq:surface_tension}
  \gamma &= f_s(c,\mathbf{P})-c\partial_c f_s(c,\mathbf{P}) - \mathbf{P}\cdot \bm{\nabla}_\mathbf{P} f_s(c,\mathbf{P}) -\frac{\kappa_c}{2}|\bm{\nabla}_\mathbf{r}c|^2 +\kappa_c c\bm{\nabla}_\mathbf{r}^2 c + \frac{a}{2} \frac{\left|\mathbf{P}\right|^2}{c^2} + \frac{b}{4} \frac{\left|\mathbf{P}\right|^4}{c^4}  \nonumber \\
  &+ \frac{\kappa_P}{2} \left(\bm{\nabla}_\mathbf{r} \frac{\mathbf{P}}{c}\right) : \left(\bm{\nabla}_\mathbf{r} \frac{\mathbf{P}}{c}\right)
  + W \frac{\mathbf{P}}{c} \cdot \hat{\mathbf{n}},
\end{align}
where the terms involving $\mathbf{P}$ are in addition to the usual form of
$\gamma$ when surfactant orientability is not
considered\cite{wilczek2015modelling}. From Eq. (\ref{eq:height_var}) we then
have that,
\begin{equation}
  \frac{\delta \mathcal{F}_I}{\delta h(\mathbf{r},t)}=\partial_h \Phi(h) - \bm{\nabla}_\mathbf{r} \cdot (\gamma \bm{\nabla}_\mathbf{r}h).
\end{equation}
This definition of the surface tension, Eq. (\ref{eq:surface_tension}), can
be understood by proving the equivalence between the (generalised) surface
tension and the (generalised) local surface grand-potential
density\cite{thiele2012thermodynamically}, i.e. $\gamma=\omega_s$, where the
surface grand potential is given as,
\begin{equation}\label{eq:grand_potential}
  \Omega_s[c,\mathbf{P}]= \int d\mathbf{r} \xi \omega_s = \int d\mathbf{r} \xi \left(g_s(c,\mathbf{P})-c \mu_s - \mathbf{P} \cdot \mathbf{O} \right)
\end{equation}
where $\mu_s$ is the surface-chemical potential and $\mathbf{O}$ is the
orientation field potential, i.e. the conjugate field to the polarisation,
$\mathbf{P}$. $g_s(c,\mathbf{P})$ is the Helmholtz free energy due to the
presence of the orientable surfactants, such that, as compared to Eq.
(\ref{eq:free_energy_functional}) the free-energy functional is written,
\begin{equation}\label{eq:free_energy_surface}
  \mathcal{F}_I[h,c,\mathbf{P}]=\int d\mathbf{r} \left[ \Phi(h)+ \xi g_s(c,\mathbf{P})\right].
\end{equation}
The surface chemical potential and orientation potential field are related to their conjugate variables via the variational derivatives,
\begin{equation}
  \mu_s=\frac{\delta \mathcal{F}_I}{\delta \widetilde{c}}, \ \text{and}, \ \mathbf{O} = \frac{\delta \mathcal{F}_I}{\delta \widetilde{\mathbf{P}}},
\end{equation}
which from inserting Eqs. (\ref{eq:concentration_derivative}) and (\ref{eq:polarisation_derivative}), along with the local surface free energy, $g_s$, defined in Eqs. (\ref{eq:free_energy_surface}) and (\ref{eq:free_energy_functional}), into Eq. (\ref{eq:grand_potential}), and from the surface tension in Eq. (\ref{eq:surface_tension}), one obtains
\begin{equation}
  \Omega_s[c,\mathbf{P}]=\int d\mathbf{r} \xi \gamma (c,\mathbf{P}),
\end{equation}
which gives the equivalence between the surface tension and local grand
potential, $\gamma=\omega_s$, as required.

The result of this equivalence means one can connect with thermodynamics and
derive a Gibbs-Duhem like equation (at constant temperature) for the gradient
of the surface tension, i.e. from Eqs. (\ref{eq:concentration_derivative}),
(\ref{eq:polarisation_derivative}) and (\ref{eq:surface_tension}) it can be
shown that,
\begin{equation}\label{eq:grad_surface_tension}
  \bm{\nabla}_\mathbf{r} \gamma = -c \bm{\nabla}_\mathbf{r} \frac{\delta \mathcal{F}_I}{\delta \widetilde{c}(\mathbf{r},t)} - \left(\bm{\mathbf{P}} \cdot \bm{\nabla}_\mathbf{r}\right)
  \frac{\delta \mathcal{F}_I}{\delta \widetilde{\mathbf{P}}(\mathbf{r},t)}.
\end{equation}
Now, from Eq. (\ref{eq:grad_surface_tension}) we are able to write the
Marangoni flow terms, $(h/\eta)c\bm{\nabla}_\mathbf{r}\gamma$ and
$(h/\eta)\mathbf{P}\otimes \bm{\nabla}_\mathbf{r}\gamma$, in Eqs.
(\ref{eq:concentration_thin_film}) and (\ref{eq:polarisation_thin_film}),
respectively in terms of the free-energy functional, meaning the set of
thin-film equations, (\ref{eq:thin_film}), (\ref{eq:concentration_thin_film})
and (\ref{eq:polarisation_thin_film}), may be represented fully in a gradient
dynamics form as,
\begin{subequations}\label{eq:thin_film_set}
\begin{align}
  \partial_t h &= \bm{\nabla}_\mathbf{r} \cdot \left[ \frac{h^3}{3\eta} \bm{\nabla}_\mathbf{r}\frac{\delta \mathcal{F}_I}{\delta h}  + \frac{h^2}{2\eta} \left(c \bm{\nabla}_\mathbf{r}
  \frac{\delta \mathcal{F}_I}{\delta \widetilde{c}} + \left(\mathbf{P} \cdot \bm{\nabla}_\mathbf{r}\right) \frac{\delta \mathcal{F}_I}{\delta \widetilde{\mathbf{P}}} \right) \right],\label{eq:grad_thin_film}\\
  \partial_t c &= \bm{\nabla}_\mathbf{r} \cdot \left[ \frac{h^2 c}{2\eta} \bm{\nabla}_\mathbf{r} \frac{\delta \mathcal{F}_I}{\delta h} + \frac{h}{\eta} c \left(c \bm{\nabla}_\mathbf{r}
  \frac{\delta \mathcal{F}_I}{\delta \widetilde{c}} + \left(\mathbf{P} \cdot \bm{\nabla}_\mathbf{r}\right) \frac{\delta \mathcal{F}_I}{\delta \widetilde{\mathbf{P}}}  \right) \right] +\bm{\nabla}_\mathbf{r}\cdot \left[(D_\parallel-D_\perp) \mathbf{Q}^{xy}\cdot \bm{\nabla}_\mathbf{r}
  \frac{\delta \mathcal{F}_I}{\delta \widetilde{c}}\right] \nonumber \\
  &+ \bm{\nabla}_\mathbf{r} \cdot \left[\left(\frac{D_\parallel}{3}+\frac{2D_\perp}{3} \right) c \bm{\nabla}_\mathbf{r} \frac{\delta \mathcal{F}_I}{\delta \widetilde{c}}\right] +\bm{\nabla}_\mathbf{r} \cdot \left[D_\perp \left(\mathbf{P} \cdot \bm{\nabla}_\mathbf{r}\right)
  \frac{\delta \mathcal{F}_I}{\delta \widetilde{\mathbf{P}}}\right],\\
  \partial_t \mathbf{P} &= \bm{\nabla}_\mathbf{r} \cdot \left[ \frac{h^2 }{2\eta} \mathbf{P} \otimes \bm{\nabla}_\mathbf{r} \frac{\delta \mathcal{F}_I}{\delta h} + \frac{h}{\eta} \mathbf{P} \otimes \left(c \bm{\nabla}_\mathbf{r}
  \frac{\delta \mathcal{F}_I}{\delta \widetilde{c}} + \left(\mathbf{P} \cdot \bm{\nabla}_\mathbf{r}\right) \frac{\delta \mathcal{F}_I}{\delta \widetilde{\mathbf{P}}} \right) \right] + \bm{\nabla}_\mathbf{r} \cdot \left[D_\perp \mathbf{P} \otimes \bm{\nabla}_\mathbf{r}
  \frac{\delta \mathcal{F}_I}{\delta \widetilde{c}}\right] \nonumber \\
  &+ \mathcal{D}_\text{rot} \left(\mathbf{Q}-\frac{2}{3}c \mathbf{I} \right)\cdot \frac{\delta \mathcal{F}_I}{\delta \widetilde{\mathbf{P}}}.
\end{align}
\end{subequations}
For the conserved quantities, $h$ and $c$, their governing equations are in a
conservative form, as is the case in non-polar surfactant molecules, or
molecules with low polarisability, whereas for the polarisation,
$\mathbf{P}$, the rotational diffusion associated with the surfactants
induces a non-conserved contribution. The Marangoni flux in Eq.
(\ref{eq:grad_thin_film}) is given in the usual
form\cite{thiele2012thermodynamically},
$-(h^2/2\eta)\bm{\nabla}_\mathbf{r}\gamma$ (from Eq.
(\ref{eq:grad_surface_tension})), however in this case, the surface tension
is of a more generalised form, Eq. (\ref{eq:surface_tension}), such that it
is dependent not only on surfactant concentration but surfactant polarisation
as well, i.e. $\gamma(c,\mathbf{P})$.

These equations are consistent with the film-height-surfactant equations in
the absence of the additional effects considered here. Indeed, if we take the
surfactants as symmetric point-like particles, the polarisation vanishes,
$D_\parallel=D_\perp=D$ and $\mathcal{D}_\text{rot}=0$, such that Eq.
(\ref{eq:thin_film_set}) reduces to the previously found set of coupled
equations for the film height and
concentration\cite{thiele2012thermodynamically}, which are given as,
\begin{subequations}
\begin{align}
  \partial_t h &= \bm{\nabla}_\mathbf{r} \cdot \left[ \frac{h^3}{3\eta} \bm{\nabla}_\mathbf{r}\frac{\delta \mathcal{F}_I}{\delta h}  + \frac{h^2 c}{2\eta}  \bm{\nabla}_\mathbf{r}
  \frac{\delta \mathcal{F}_I}{\delta \widetilde{c}}  \right],\\
  \partial_t c &= \bm{\nabla}_\mathbf{r} \cdot \left[ \frac{h^2 c}{2\eta} \bm{\nabla}_\mathbf{r} \frac{\delta \mathcal{F}_I}{\delta h} + \frac{h c^2}{\eta}  \bm{\nabla}_\mathbf{r}
  \frac{\delta \mathcal{F}_I}{\delta \widetilde{c}} \right] + \bm{\nabla}_\mathbf{r} \cdot \left[D c \bm{\nabla}_\mathbf{r} \frac{\delta \mathcal{F}_I}{\delta \widetilde{c}}\right].
\end{align}
\end{subequations}

%\section{Discussion \& Conclusions}

\section{Summary and Conclusions}

We have considered the case of a surfactant-laden thin liquid film, with the
surfactant molecules being treated as anisotropic, such that they have a
head-tail polar structure and diffuse across the film surface as well as
rotationally. We have presented a microscopic derivation of the interface
transport equation for such molecules, using elements from the statistical
mechanics of classical fluids, namely DFT, and its dynamic extension, DDFT.

First, we introduced the general DDFT for uniaxial orientable colloids and
generalised this to the case in which this dynamics is occurring on the
surface of a thin film. Invoking the long-wave approximation we are able to
find the equation governing the joint surfactant concentration and
orientation density, which is coupled to the hydrodynamic equation for the
film height that includes the gradient of interfacial tension. We take that
the only contributions to the free-energy functional associated with the
surfactant dynamics are through the surfactant concentration and polarisation
field, which are appropriately defined as the zeroth- and first-moment of the
joint distribution with respect to the orientation. We then take the
respective moments of the governing equation to find thin-film equations for
the surfactant concentration and polarisation field. To fully close this set
of equations, we express the free energy-functional in its explicit form in
terms of the film height, surfactant concentration and polarisation, where
there are contributions from the disjoining pressure, the free energy of the
surfactants, as well as terms related to the polarisation of the surfactants.
After computing the variations of the free-energy functional with respect to
the fields of interest, and defining a generalised surface tension, we
finally arrive at the set of equations for a thin film laden with orientable
polar uniaxial surfactants. These equations, are of gradient dynamics form,
as they should be, and they reduce to the usual equations for nonpolar
surfactants.

There are a number of interesting questions related to the analysis presented
here, and further avenues for which the results may be explored. Firstly, it
would be interesting to investigate how to close the set of thin-film
equations, (\ref{eq:thin_film_set}), via the approximation of the
$\mathbf{Q}$ alignment tensor through the polarisation, $\mathbf{P}$, where
existing techniques from liquid crystal theory may be
useful~\cite{de1993physics}. Alternatively, one could truncate the moment
hierarchy at a higher order, i.e. derive an additional equation for
$\mathbf{Q}$, and appropriately devise closure approximations for the higher
moments. It would also be interesting to allow the surfactants to be soluble
such that an extra observable, the dissolved surfactant density, would need
to be included in the formulation. In this case, the orientation of the
surfactants would have a significant impact on surface adsorption; in
particular, the adsorption flux~\cite{thiele2016gradient} can be taken to be
dependent on the polarisation, leading to a more realistic model for thin
films with soluble surfactants.

The utility of the gradient form of the thin-film equations allows one to
simply extend the free-energy functional to more accurate approximations,
namely, weighted-density approximations, fundamental measure theory,
statistical associating fluid theory and molecular DFT, e.g. Refs.
\citenum{evans1979nature,tarazona1985free,rosenfeld1989free,wu2006density,Yatsyshin2012,jeanmairet2013molecular,Andreas2014},
as well as including higher-order polarisation induced contributions to the
free-energy
functional~\cite{marchetti2013hydrodynamics,trinschek2020thin,stegemerten2022symmetry}.
Furthermore, with recent advances in the application of statistical-learning
techniques to
DFT~\cite{yatsyshin2022physics,sammuller2023neural,sammuller2025neural,simon2026machine}
and beyond~\cite{Antonio2024,Antonio2025}, it would be interesting to apply
these to gain a greater understanding of the effect of surfactant orientation
on the dynamics. In particular, it would be interesting to use molecular
dynamics of orientable surfactants to try and glean the structure of the
free-energy functional via functional mappings which would go beyond the
adiabatic approximation\cite{de2023perspective,zimmermann2024neural}.

On the hydrodynamics front, the majority of thin-film studies are typically
based on macroscopic models, as done here. There are some notable exceptions,
e.g. the recent work in Ref.~\citenum{teVrugt2024microscopic} which offered a
microscopic derivation of the thin-film equation using the Mori-Zwanzig
projection-operator formalism. Of particular interest here, would be a fully
microscopic derivation, where in addition to the surfactant interfacial
transport equation, we also have a microscopic derivation of the hydrodynamic
equation for the film height which involves the gradient of interfacial
tension. Extending these equations to allow for stochasticity-noise would
enable scrutiny of the effects of thermal fluctuations,
e.g.~Ref.~\citenum{duran2019instability}. Further, one may include
non-conserved dynamics such as
evaporation/condensation\cite{thiele2016gradient} or multiple species of
reactive surfactants\cite{voss2024gradient}, which could be taken to be
dependent on the surfactant polarisation; or indeed additional levels of
complexity such as reactive surfactants~\cite{Antonio2007}, where the
coupling of thin-film hydrodynamics and surfactant chemistry can either
stabilise instabilities occurring in the pure chemical subsystem, or indeed
it can induce instabilities in regions where the pure hydrodynamic and
chemical subsystems are both stable. We shall examine these and related
problems in future studies.

\section*{Acknowledgements}

We are grateful to the anonymous reviewers for valuable comments and critical
suggestions. We acknowledge financial support by the ERC-EPSRC Frontier
Research Guarantee through Grant No. EP/X038645, ERC through Advanced Grant
No. 247031, and EPSRC through Grant Nos. EP/L025159 and EP/L020564.

\section{Appendix}

\subsection{A. Derivation of Eq. (\ref{eq:surface_DDFT})}

We show how we derive the surface DDFT equation (\ref{eq:surface_DDFT}) from
Eq. (\ref{eq:DDFT}). We start by considering the continuity equation for the
conserved surfactant concentration diffusing on the free surface, $S(t)$,
\begin{equation}
  \frac{d}{dt}\int_{S(t)} dS \int_\Omega d\hat{\bm{\omega}} \Gamma(\mathbf{r},\hat{\bm{\omega}},t) +  \int_{\partial S(t)} d\ell_s \int_\Omega d\hat{\bm{\omega}} \mathbf{J}(\mathbf{r},\hat{\bm{\omega}},t) \cdot \hat{\mathbf{n}}_s + \int_{S(t)}dS \int_{\partial \Omega} d\ell_{\hat{\bm{\omega}}} \bm{\mathcal{J}}(\mathbf{r},\hat{\bm{\omega}},t) \cdot \hat{\mathbf{n}}_\Omega=0,
\end{equation}
where $d/dt=\partial/\partial t + \mathbf{v}_s \cdot \bm{\nabla}_s$ is the
material derivative (see comments below Eq. (\ref{eq:normal}) and
Ref.~\citenum{Antonio2007} about the subtleties of this material derivative).
$\mathbf{J}(\mathbf{r},\hat{\bm{\omega}},t)$ and
$\bm{\mathcal{J}}(\mathbf{r},\hat{\bm{\omega}},t)$ are the fluxes associated
with the translational diffusion and rotational diffusion, respectively.
$\Omega$ is an enclosed region of orientation space, i.e a subset of the unit
sphere, and $d\ell_s$ and $d\ell_{\hat{\bm{\omega}}}$ are line elements on
the surface and unit sphere, respectively. $\hat{\mathbf{n}}_s$ and
$\hat{\mathbf{n}}_\Omega$ are the unit normal vectors to $S$ and $\Omega$,
respectively. Using the divergence theorem for surfaces and spheres we have
\begin{equation}
  \frac{d}{dt}\int_{S(t)} dS \int_\Omega d\hat{\bm{\omega}} \Gamma(\mathbf{r},\hat{\bm{\omega}},t) +  \int_{S(t)} dS \int_\Omega d\hat{\bm{\omega}} \left[\bm{\nabla}_s \cdot \mathbf{J}(\mathbf{r},\hat{\bm{\omega}},t) +  \bm{\nabla}_{\hat{\bm{\omega}}} \cdot \bm{\mathcal{J}}(\mathbf{r},\hat{\bm{\omega}},t)\right]=0.
\end{equation}
Now, we may bring the material derivative inside the integral, but as the surface is evolving we must use the product rule along with the identity\cite{batchelor1967introduction},
\begin{equation}
  \frac{d}{dt} dS = dS \bm{\nabla}_s \cdot \mathbf{v},
\end{equation}
to obtain,
\begin{equation}\label{eq:continuity_integral}
  \int_{S(t)} dS \int_\Omega d\hat{\bm{\omega}} \left[\frac{\partial}{\partial t} \Gamma + \bm{\nabla}_s \cdot \left( \mathbf{v}\Gamma\right)\right] +  \int_{S(t)} dS \int_\Omega d\hat{\bm{\omega}} \left[\bm{\nabla}_s \cdot \mathbf{J}(\mathbf{r},\hat{\bm{\omega}},t) +  \bm{\nabla}_{\hat{\bm{\omega}}} \cdot \bm{\mathcal{J}}(\mathbf{r},\hat{\bm{\omega}},t)\right]=0.
\end{equation}
By writing Eq. (\ref{eq:DDFT}) in the form of a continuity equation,
\begin{equation}
  \partial_t \rho(\bm{\mathdutchcal{r}},\hat{\bm{\omega}},t)= -\bm{\nabla}_{\bm{\mathdutchcal{r}}} \cdot \mathbf{J}(\bm{\mathdutchcal{r}},\hat{\bm{\omega}},t) - \bm{\nabla}_{\hat{\bm{\omega}}} \cdot \bm{\mathcal{J}}(\bm{\mathdutchcal{r}},\hat{\bm{\omega}},t)
\end{equation}
we are able to glean the expressions for the fluxes as,
\begin{subequations}
  \begin{align}
  \mathbf{J}(\bm{\mathdutchcal{r}},\hat{\bm{\omega}},t)&=-\mathbf{D}(\hat{\bm{\omega}}) \cdot \rho(\bm{\mathdutchcal{r}},\hat{\bm{\omega}},t) \bm{\nabla}_{\bm{\mathdutchcal{r}}} \frac{\delta \mathcal{F}}{\delta \rho} \\
  \bm{\mathcal{J}}(\bm{\mathdutchcal{r}},\hat{\bm{\omega}},t)&= \mathcal{D}_\text{rot} \hat{\bm{\omega}} \times \left(\rho(\bm{\mathdutchcal{r}},\hat{\bm{\omega}},t) \hat{\bm{\omega}} \times \bm{\nabla}_{\hat{\bm{\omega}}} \frac{\delta \mathcal{F}}{\delta \rho}\right),
\end{align}
\end{subequations}
where we have used $\left(\hat{\bm{\omega}}\times
\bm{\nabla}_{\hat{\bm{\omega}}}\right) \cdot \left( \rho \hat{\bm{\omega}}
\times \bm{\nabla}_{\hat{\bm{\omega}}} \delta \mathcal{F}/\delta \rho
\right)=-\bm{\nabla}_{\hat{\bm{\omega}}} \cdot \left[\hat{\bm{\omega}} \times
\left(\rho \hat{\bm{\omega}} \times \bm{\nabla}_{\hat{\bm{\omega}}} \delta
\mathcal{F}/\delta \rho \right)\right]$. Now, by constraining these fluxes to
the surface $z=h(\mathbf{r},t)$, we have
$\mathbf{J}(\bm{\mathdutchcal{r}},\hat{\bm{\omega}},t) \rightarrow
\mathbf{J}(\mathbf{r},\hat{\bm{\omega}},t)$,
$\bm{\mathcal{J}}(\bm{\mathdutchcal{r}},\hat{\bm{\omega}},t) \rightarrow
\bm{\mathcal{J}}(\mathbf{r},\hat{\bm{\omega}},t)$ and
$\bm{\nabla}_{\bm{\mathdutchcal{r}}} \rightarrow \bm{\nabla}_s$, then writing
in terms of the concentration, $\Gamma(\mathbf{r},\hat{\bm{\omega}},t)$, and
inserting into Eq. (\ref{eq:continuity_integral}), we find
\begin{align}
  \frac{\partial}{\partial t} \Gamma + \bm{\nabla}_s \cdot \left( \mathbf{v}\Gamma\right) &=  \bm{\nabla}_s \cdot \left[\mathbf{D}(\hat{\bm{\omega}}) \cdot \Gamma  \bm{\nabla}_s \frac{\delta \mathcal{F}}{\delta \Gamma} \right] - \mathcal{D}_\text{rot} \bm{\nabla}_{\hat{\bm{\omega}}} \cdot \left[\hat{\bm{\omega}} \times \left(\Gamma \hat{\bm{\omega}} \times \bm{\nabla}_{\hat{\bm{\omega}}} \frac{\delta \mathcal{F}}{\delta \Gamma}\right) \right] \nonumber\\
  &=\bm{\nabla}_s \cdot \left[\mathbf{D}(\hat{\bm{\omega}}) \cdot \Gamma  \bm{\nabla}_s \frac{\delta \mathcal{F}}{\delta \Gamma} \right] + \mathcal{D}_\text{rot} \hat{\bm{\omega}} \times \bm{\nabla}_{\hat{\bm{\omega}}} \cdot \left[\Gamma \hat{\bm{\omega}} \times \bm{\nabla}_{\hat{\bm{\omega}}} \frac{\delta \mathcal{F}}{\delta \Gamma} \right],
\end{align}
where we have neglected the integrals as $\Gamma$ is conserved and the
surface and orientation is arbitrary. Finally, by writing $\mathbf{v}$ in
terms of its tangential, $\mathbf{v}_s=(\mathbf{I}-\hat{\mathbf{n}}\otimes
\hat{\mathbf{n}})\cdot \mathbf{v}$, and normal,
$\mathbf{v}_n=(\mathbf{v}\cdot \hat{\mathbf{n}})\hat{\mathbf{n}}$,
components, we obtain Eq. (\ref{eq:surface_DDFT}).

\subsection{B. Calculation of Rotational Term in Eq. (\ref{eq:polarisation_thin_film})}

Here we illustrate how taking the average over the final term on the right
hand side (RHS) of Eq. (\ref{eq:joint_concentration_thin_film}) with respect
to $\hat{\bm{\omega}}$, via Eq. (\ref{eq:polarisation}), leads to the final
term on the RHS of Eq. (\ref{eq:polarisation_thin_film}). Inserting Eq.
(\ref{eq:chain_rule}) into Eq. (\ref{eq:joint_concentration_thin_film}) leads
to the expression for the final term as,
\begin{equation}\label{eq:final_term}
  \mathcal{D}_\text{rot} \hat{\bm{\omega}} \times \bm{\nabla}_{\hat{\bm{\omega}}} \cdot \left[\Gamma \hat{\bm{\omega}} \times \bm{\nabla}_{\hat{\bm{\omega}}}
  \frac{\delta \mathcal{F}_I}{\delta \widetilde{\Gamma}} \right]=\mathcal{D}_\text{rot} \hat{\bm{\omega}} \times \bm{\nabla}_{\hat{\bm{\omega}}} \cdot \left[\Gamma \hat{\bm{\omega}} \times
  \frac{\delta \mathcal{F}_I}{\delta \widetilde{\mathbf{P}}} \right].
\end{equation}
Averaging over Eq.~(\ref{eq:final_term}) with respect to $\hat{\bm{\omega}}$
gives,
\begin{align}
  &\mathcal{D}_\text{rot} \oint d\hat{\bm{\omega}} \ \hat{\bm{\omega}} \left(\hat{\bm{\omega}} \times \bm{\nabla}_{\hat{\bm{\omega}}}\right) \cdot \left[\Gamma \hat{\bm{\omega}} \times
  \frac{\delta \mathcal{F}_I}{\delta \widetilde{\mathbf{P}}} \right]=\mathcal{D}_\text{rot} \oint d\hat{\bm{\omega}} \ \hat{\bm{\omega}} \left[\hat{\bm{\omega}} \cdot \left(\bm{\nabla}_{\hat{\bm{\omega}}} \times \left( \Gamma \hat{\bm{\omega}} \times
  \frac{\delta \mathcal{F}_I}{\delta \widetilde{\mathbf{P}}}\right)\right) \right] \nonumber \\
  &=-\mathcal{D}_\text{rot} \oint d\hat{\bm{\omega}} \ \hat{\bm{\omega}} \left[\bm{\nabla}_{\hat{\bm{\omega}}} \cdot \left(\hat{\bm{\omega}} \times \left( \Gamma \hat{\bm{\omega}} \times
  \frac{\delta \mathcal{F}_I}{\delta \widetilde{\mathbf{P}}}\right)\right) \right],
\end{align}
since $\bm{\nabla}_{\hat{\bm{\omega}}}\times \hat{\bm{\omega}}=\mathbf{0}$. Now, by integrating by parts we obtain,
\begin{align}
  &-\mathcal{D}_\text{rot} \oint d\hat{\bm{\omega}} \ \hat{\bm{\omega}} \left[\bm{\nabla}_{\hat{\bm{\omega}}} \cdot \left(\hat{\bm{\omega}} \times \left( \Gamma \hat{\bm{\omega}} \times
  \frac{\delta \mathcal{F}_I}{\delta \widetilde{\mathbf{P}}}\right)\right) \right]=\mathcal{D}_\text{rot} \oint d\hat{\bm{\omega}} \ \left(\bm{\nabla}_{\hat{\bm{\omega}}}\hat{\bm{\omega}}\right) \cdot \left(\hat{\bm{\omega}} \times \left( \Gamma \hat{\bm{\omega}} \times
  \frac{\delta \mathcal{F}_I}{\delta \widetilde{\mathbf{P}}}\right)\right) \nonumber\\
  &=\mathcal{D}_\text{rot} \oint d\hat{\bm{\omega}} \left( \mathbf{I}-\hat{\bm{\omega}}\otimes \hat{\bm{\omega}}\right) \cdot \left[ \hat{\bm{\omega}}\left( \Gamma \hat{\bm{\omega}} \cdot
  \frac{\delta \mathcal{F}_I}{\delta \widetilde{\mathbf{P}}} \right) -\Gamma\frac{\delta \mathcal{F}_I}{\delta \widetilde{\mathbf{P}}} \right]
\end{align}
where we have used that
$\bm{\nabla}_{\hat{\bm{\omega}}}\hat{\bm{\omega}}=\mathbf{I}-\hat{\bm{\omega}}\otimes
\hat{\bm{\omega}}$ as $\bm{\nabla}_{\hat{\bm{\omega}}}$ is the gradient on
the unit sphere. We have also made use of the triple cross product expansion
along with the fact that $\hat{\bm{\omega}}$ is a unit vector.

It can be shown that the first term vanishes, i.e.,
\begin{align}
  \left( \mathbf{I}-\hat{\bm{\omega}}\otimes \hat{\bm{\omega}}\right) \cdot  \hat{\bm{\omega}}\left( \Gamma \hat{\bm{\omega}} \cdot \frac{\delta
    \mathcal{F}_I}{\delta \widetilde{\mathbf{P}}} \right)= \hat{\bm{\omega}}\left( \Gamma \hat{\bm{\omega}} \cdot
    \frac{\delta \mathcal{F}_I}{\delta \widetilde{\mathbf{P}}} \right) - \hat{\bm{\omega}} \left( \hat{\bm{\omega}} \cdot \hat{\bm{\omega}} \left( \Gamma \hat{\bm{\omega}} \cdot
    \frac{\delta \mathcal{F}_I}{\delta \widetilde{\mathbf{P}}}\right)\right)=0,
\end{align}
and thus we find
\begin{align}
  &\mathcal{D}_\text{rot} \oint d\hat{\bm{\omega}} \left( \mathbf{I}-\hat{\bm{\omega}}\otimes \hat{\bm{\omega}}\right) \cdot \left[ \hat{\bm{\omega}}\left( \Gamma \hat{\bm{\omega}} \cdot
  \frac{\delta \mathcal{F}_I}{\delta \widetilde{\mathbf{P}}} \right) -\Gamma\frac{\delta \mathcal{F}_I}{\delta \widetilde{\mathbf{P}}} \right]= - \mathcal{D}_\text{rot} \oint d\hat{\bm{\omega}} \left( \mathbf{I}-\hat{\bm{\omega}}\otimes \hat{\bm{\omega}}\right) \cdot
  \Gamma\frac{\delta \mathcal{F}_I}{\delta \widetilde{\mathbf{P}}} \nonumber \\
  &=\mathcal{D}_\text{rot} \left(\mathbf{Q}-\frac{2}{3}c \right) \frac{\delta \mathcal{F}_I}{\delta \widetilde{\mathbf{P}}},
\end{align}
where we have utilised Eq.~(\ref{eq:Q_tensor}), which gives exactly the final
term on the RHS of Eq.~(\ref{eq:polarisation_thin_film}).

\bibliography{References}

\newpage

% \begin{figure}
%   \centering
%   \includegraphics[width = 0.75\textwidth]{TOC Graphic.pdf}
%   \caption{TOC Graphic}
% \end{figure}

\end{document}